\documentclass[amsmath,showpacs,twocolumn,superscriptaddress,aps,prl]{revtex4}
\usepackage{bm}
\usepackage{graphics}
\usepackage{graphicx}

\newcommand{\bq}{\begin{equation}}
\newcommand{\ee}{\end{equation}}
\newcommand{\fr}[2]{\frac{#1}{#2}}

\begin{document}

\title{Guided plasmons in graphene $p$-$n$  junctions}
\author{E.~G.~Mishchenko}
\affiliation{Department of Physics and Astronomy, University of
Utah, Salt Lake City, Utah 84112, USA}
\author{ A.~V.~Shytov$^*$}
\affiliation{Department of Physics and Astronomy, University of
Utah, Salt Lake City, Utah 84112, USA}
\author{P. G. Silvestrov}
 \affiliation{Theoretische
Physik III, Ruhr-Universit{\"a}t Bochum, 44780 Bochum, Germany}

\begin{abstract}
Spatial separation of electrons and holes in graphene gives rise to
existence of plasmon waves confined to the boundary region. Theory
of such guided plasmon modes within hydrodynamics of electron-hole
liquid is developed. For plasmon wavelengths smaller than the size
of charged domains plasmon dispersion is found to be $\omega \propto
q^{1/4}$. Frequency, velocity and direction of propagation of guided
plasmon modes can be easily controlled by external electric field.
In the presence of magnetic field spectrum of additional gapless
magnetoplasmon excitations is obtained. Our findings indicate that
graphene is a promising material for nanoplasmonics.
\end{abstract}


\maketitle

{\it Introduction}. Breakthrough progress in  synthesis and
characterization  has made graphene \cite{Wil} a promising object
for nanoelectronics. Operation of graphene-based transistors
\cite{GN} and other components would rely on the properties of its
{\it single-particle} excitations -- electrons and holes. However,
one can also envisage a completely different set of applications
which employ {\it collective} excitations, such as plasmons.
Currently, plasmon excitations in metallic structures are a subject
of nanoplasmonics, a new field which has emerged at the confluence
of optics and condensed matter physics with one of the aims being
the developing of plasmon-enhanced high resolution near-field
imaging methods \cite{HAA,SAM}. Another objective is possible
utilization of plasmons in integrated optical circuits. However,
perspectives of graphene for nanoplasmonics are largely unexplored
since plasmon modes of graphene flakes have not been addressed so
far. As our results indicate a great amount of control over graphene
plasmon properties makes it a very promising material for
applications.

Fundamentally, the spectrum of  collective charge oscillations
reflects the long-range nature of Coulomb  interaction. In
conventional two dimensional systems, such as those created in
semiconducting heterostructures, plasmons are gapless,
$\omega^2(q)=2\pi e^2 n q/m^*$, with $n$ and $m^*$ being electron
density and effective mass, respectively \cite{Stern}. Such
oscillations can be treated hydrodynamically.
In clean graphene at zero temperature the plasmon frequency,
$\omega^2 \propto |E_F|$, vanishes with decreasing the doping level
$E_F$. It has been argued \cite{SFM} that the interaction between
electrons and holes in the final state can modify the response
functions of Dirac fermions and open up a possibility for the
propagation of charge oscillations at low frequencies $\omega < qv$,
where $v$ is electron velocity. Still, hydrodynamic ($\omega > qv$)
analog of conventional plasmons remains absent unless either
temperature is non-zero \cite{T_plasmon} or graphene is driven away
from the charge neutrality point by doping or gating
\cite{E_plasmon}. Expectedly, in both cases plasmon spectrum has the
conventional form, $\omega(q) \propto q^{1/2}$.

\begin{figure}[h]
\centerline{\includegraphics[width=85mm,angle=0]{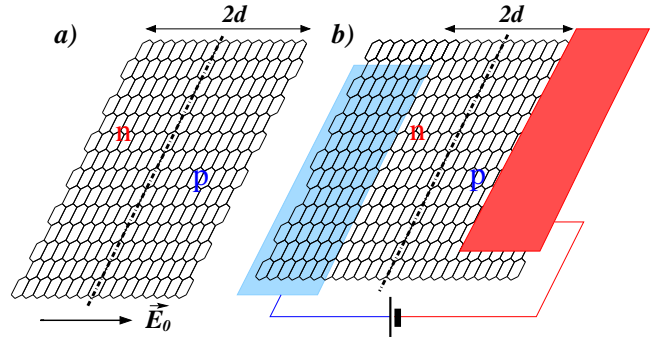}}
\caption{Two types of graphene $p$-$n$ junctions: a) field-induced,
b) gate-induced. Dot-dashed line indicates boundary between electron
and hole regions and, correspondingly, the direction of plasmon
propagation. In case of field-induced junction it is controlled by
the direction of external electric field $E_0$.} \label{fig1}
\end{figure}

In the present paper we investigate  spectra of hydrodynamic
plasmons in spatially inhomogeneous graphene flakes. Realistic
graphene samples are typically subject to disorder potential and
mechanical strain \cite{cn} that lead to the formation of charged
electron and hole puddles \cite{puddles} with boundaries between $n$
and $p$ regions being the lines of zero chemical potential.
Moreover, controlled $p$-$n$ junctions can be made with the help of
metallic gates \cite{Mar}. Also $p$-$n$ junctions can be created by
applying electric field within the plane of a graphene flake, see
Fig.~\ref{fig1}a. The field separates electrons and holes spatially
in a way that allows control of both the amount of induced charge
(and thus plasmon frequency) and spatial orientation of the junction
(the direction of plasmon propagation).

Below, we demonstrate that such $p$-$n$ junctions can guide
plasmons. We show the existence of charge oscillations which are
localized at the junction and have the amplitude decaying with the
distance to the junction. For wavelengths shorter than the width of
the charged domains, we find the plasmon spectrum of the form,
\begin{equation}
\label{result} \omega_n^2(q) =\alpha_n \frac{e^2
v}{\hbar}\sqrt{\frac{q |\rho'_0|}{e}},
\end{equation}
where $\rho'_0$ is the gradient of equilibrium charge density at the
junction, $v$ is electron velocity, and $n=0,1,2,...$ enumerates the
solutions. The lowest mode has
$\alpha_0=4\sqrt{2\pi}\Gamma(3/4)/\Gamma(1/4) \approx 3.39$.

 Below we
derive this result and discuss plasmon properties for the two types
of $p$-$n$ junctions: electric field controlled and gate controlled,
as shown in Fig.~1.

{\it Hydrodynamics of charge density oscillations.} We utilize the
hydrodynamic approach to describe the motion of charged Dirac
fermions. The rate of change of electric current density ${\bf J}$
due to dynamic electric field ${\bf E}$ follows from the usual
intra-band Drude conductivity with the corresponding density of
states \cite{LLV},
\begin{equation}
\label{motion} {\bf \dot J}({\bf r},t)
=\frac{e^2}{\pi \hbar^2} |\mu({\bf r})| {\bf E}({\bf r},t),
\end{equation}
determined by the local value of  chemical potential $\mu({\bf r})$
as measured from the Dirac point (positive for electrons and
negative for holes).
 Electric
current is related to the variation of charge density
$\delta \rho$ by means of the continuity equation,
\begin{equation}
\label{continuity} \delta \dot \rho ({\bf r},t) + \nabla \cdot {\bf J}({\bf r},t)=0.
\end{equation}
Finally, the variation of charge density produces electric field
according to the Coulomb law \cite{ftn1},
\begin{equation}
\label{coulomb} {\bf E}({\bf r},t) =-\nabla \int d^2r' \frac{\delta
\rho ({\bf r'},t)}{|{\bf r}-{\bf r'}|}.
\end{equation}
Equations (\ref{motion})-(\ref{coulomb}) give a closed system for
plasmon excitations in graphene flakes. We apply it to a $p$-$n$
junction created in a strip infinite along the $y$-axis (direction
of plasmon propagation). Using the Fourier representation, $\delta
\rho ({\bf r},t) =\delta \rho(x) \exp(iqy-i\omega t)$, and
eliminating ${\bf E}$ and ${\bf J}$ we arrive at the equation for
the oscillating part of electron density,
\begin{eqnarray}
\label{integro-diff} \omega^2\delta
\rho(x)+\frac{2e^2v}{\sqrt{\pi} \hbar} \left\{\frac{d}{d x}
\sqrt{\frac{|\rho_0(x)|}{e}} \frac{d}{d x} -q^2 \sqrt{\frac{|\rho_0(x)|}{e}} \right\} \nonumber \\
\times \int_{-d}^d d x' \delta \rho(x') K_0(|q||x-x'|)=0,
\end{eqnarray}
Here $K_0$ is the modified Bessel function and $2d$ is the width of
graphene flake. Within the Thomas-Fermi approximation equilibrium
charge density $\rho_0(x)$ is related to the chemical potential via
$\rho_0(x)=-\text{sgn}(\mu) e\mu^2(x)/\pi\hbar^2 v^2$ (electron
charge is taken to be $-e$). This follows from the condition that
the electrochemical potential $\mu(x)-e\phi(x)$ is constant
throughout the system. The solutions of Eq.~(\ref{integro-diff})
will now be considered  for large and small plasmon momenta
separately.

{\it Short wavelength, $q \gg 1/d$}. In this case the decay of
plasmon density $\delta \rho (x)$ occurs over a distance much
smaller than the width of the system and the limits of integration
in Eq.~(\ref{integro-diff}) can be extended to infinity. Assuming
(cf.\ Eq.~(\ref{approx_density}) below) the linear dependence,
$\rho_0(x) =\rho'_0 x$, we observe that the integro-differential
equation (\ref{integro-diff}) acquires obvious scaling property.
Introducing the variable $\xi=qx$ we arrive at the plasmon spectrum
in the form (\ref{result}), with dimensionless constants $\alpha_n$
determined from the eigenvalue problem:
 \begin{eqnarray}\label{eq7}
-\frac{2}{\sqrt{\pi}} \left(\frac{d}{d\xi} \sqrt{|\xi|}
\frac{d}{d\xi} - \sqrt{|\xi|} \right) \nonumber \\ \times
\int_{-\infty}^\infty d\xi' \delta \rho^{(n)}(\xi')
K_0(|\xi-\xi'|)=\alpha_n \delta \rho^{(n)}(\xi).
 \end{eqnarray}
Interestingly, this integro-differential equation allows a complete
analytic solution, though the detailed analysis is beyond the scope
of this paper. Our main findings are as follows.
Solutions are enumerated by $n=0,1,2...$ with even/odd numbers
corresponding to even/odd density profile, $\delta\rho^{(n)}(-
\xi)=(-1)^n \delta\rho^{(n)}(\xi)$. Surprisingly, eigenvalues are
{\it doubly-degenerate} and given by
 \bq
 \alpha_{2n}=\alpha_0
\fr{2n+1}{4n+1}\cdot \fr{3\cdot 7\cdot\cdot (4n-1)}{1\cdot
5\cdot\cdot (4n-3)}, ~~~\alpha_{2n+1}= \alpha_{2n}.
 \ee
At large distances all modes have exponential dependence,
$\delta\rho^{(n)}(\xi) \sim e^{-|\xi|}$, while at $|\xi|\ll 1$  even
and odd solutions exhibit different behavior, $\delta\rho^{({\rm
even})}\sim 1- {\rm const}\sqrt{|\xi|}$ and $\delta\rho^{({\rm
odd})}\sim {{\rm sign}(\xi)}/{\sqrt{|\xi|}}$. The first pair of
solutions (belonging to the lowest eigenvalue $\alpha_0$) in the
Fourier representation $\delta \rho^{(n)} (k)=\int d\xi
\delta\rho^{(n)}(\xi)e^{ik\xi}$ acquires a simple
form:
 \bq
 \label{low_pair}
\delta \rho^{(0)}(k)\propto \fr{1}{(1+k^2)^{3/4}}, ~~~ \delta
\rho^{(1)}(k) \propto \fr{k}{(1+k^2)^{3/4}}.
 \ee



{\it Long wavelength, $q \ll 1/d$}. In contrast to the above result
(\ref{result}) plasmon spectrum at small $q$ is sensitive to a
specific realization of the $p$-$n$ junction. We address the
long-wavelength behavior of plasmons in field controlled junctions.
We expect this case to be of more interest, in addition it allows a
more complete description. Before analyzing plasmons in this
structure, we discuss the equilibrium density profile. As shown in
Fig.~1a the flake of width $2d$ is placed in external electric field
$E_0$ applied along the $x$-direction. The equilibrium density
distribution $\rho(x)$ is found from,
 \begin{equation}
\label{tf} E_0x+\text{sgn}(x)\frac{\hbar
v}{e}\sqrt{\frac{\pi}{e}|\rho_0(x)| }+ 2 \int_0^d dx' \rho_0(x')
\ln{\frac{x+x'}{|x-x'|}}=0,
 \end{equation}
where it is used that $\rho_0(x)=-\rho_0(-x)$. Prior to solving
Eq.~(\ref{tf}) it is instructive to analyze validity of the
semiclassical approach. The first condition implies that the change
of the electron wavelength is smooth on the scale of itself, $ d/dx
(\hbar v/\mu)\ll 1 $. Estimating $\mu(x) \sim eE_0x$ we obtain that
the distance to the $p$-$n$ junction line ($x=0$) should exceed the
characteristic electric field length $l_E=\sqrt{e/E_0} \ll x$. The
second condition requires that the electron wavelength is small
compared with the width of the system, $d \gg \hbar v/\mu$. Noting
that in  graphene $ \hbar v \sim e^2$  we can rewrite this second
condition simply as $l_E \ll d$. Thus, the Thomas-Fermi equation
(\ref{tf}) for the equilibrium charge density and the hydrodynamic
equation (\ref{integro-diff}) for its variation
 are applicable as long as
 \begin{equation}
\label{conditions} l_E \ll d, ~~~~ q \ll 1/l_E.
 \end{equation}
However, the ratio of $q$ and $1/d$ can be arbitrary. For a moderate
external electric field $\sim 10^4$$\text{V}$/$\text{m}$ the value
of electric length $l_E\sim 0.4\mu$$\text{m}$ and the first of the
conditions (\ref{conditions}) is satisfied easily for micron-sized
samples.

Analytic solution of Eq.~(\ref{tf}) is possible when the second term
is small, in which case the charge density is \cite{SMR}
 \begin{equation}
\label{approx_density} \rho_0(x) =\frac{E_0x}{\sqrt{d^2-x^2}}.
 \end{equation}
Substituting this expression back into Eq.~(\ref{tf}) we observe
that the second term is indeed negligible as long as $x\gg l_E^2/d$.
This is assured whenever the conditions (\ref{conditions}) are
satisfied. It is also worth pointing out that
Eq.~(\ref{approx_density}) justifies the linear approximation for
the charge density used in deriving Eq.~(\ref{result}) for $q \gg
1/d$, with $\rho'_0/e=1/(l_E^2d)$.

We now turn to the analysis of plasma oscillations propagating on
top of the density distribution, Eq.~(\ref{approx_density}). For
small plasmon momenta, $q\ll 1/d$, electric field extends beyond the
width of the flake and the equation (\ref{integro-diff}) needs to be
supplemented with the boundary condition, which ensures that
electric field (and thus the current) vanishes  at the edges, $x=\pm
d$:
 \begin{equation}
\text{P}\label{boundary} \int_{-d}^d d x~ \frac{\delta \rho(x)
}{x\pm d}=0.
 \end{equation}
 The spectrum of the lowest
symmetric mode can be most easily found by integrating
Eq.~(\ref{integro-diff}) across the width of the flake. The first
term in the brackets will then vanish exactly due to the boundary
condition (\ref{boundary}). The remaining integral can now be
calculated to the logarithmic accuracy with the help of the
approximation $K_0(q|x-x'|)=-\ln{ q|x-x'|}$:
 \begin{equation}
\label{lorarith} \int_{-d}^d d x \sqrt{\frac{|\rho_0 (x)|}{e}}
\ln{(q|x-x'|)} \approx \frac{2d \Gamma^2 (3/4)}{l_E\sqrt{\pi}}
\ln{(qd)}.
 \end{equation}
Eqs.~(\ref{integro-diff}) and (\ref{lorarith}) combine to give the
equation, $[\omega^2-\omega_0^2(q)]\int_{-d}^d dx \delta \rho(x)
=0$, that yields the dispersion of the gapless symmetric plasmon,
 \begin{equation}\label{gapless}
\omega^2_0 (q)=\Gamma^2 (3/4) \frac{4 e^2 v d}{\pi \hbar l_E} q^2
\ln{(1/qd)},
 \end{equation}
reminiscent of the plasmon spectrum in quasi-one-dimensional wires,
 The remaining modes, $n\ge 1$, are
 gapped. For these  modes  $\int_{-d}^d dx \delta \rho(x) =0$ and
simple procedure of integrating Eq.~(\ref{integro-diff}) over the
width of the flake is not useful. Instead, the equation for the
$n$-th frequency gap can be obtained by setting $q=0$ in
Eq.~(\ref{integro-diff}). We observe that
 \begin{equation}
\label{gaps} \omega^2_n (0)= \beta_n \frac{e^2 v }{ \hbar l_E d},
 \end{equation}
where $\beta_n$ are the eigenvalues of the equation,
 \begin{eqnarray}\label{gagaps}
\frac{2}{\sqrt{\pi}} \frac{d}{d\xi} \frac
{\sqrt{|\xi|}}{(1-\xi^2)^{1/4}} \int_{-1}^1 d\xi' \frac{ \delta
\rho^{(n)} (\xi')}{\xi-\xi'}= \beta_n \delta \rho^{(n)} (\xi).
 \end{eqnarray}
The zeroth mode $\beta_0 =0$, see Eq.~(\ref{gapless}), is found
analytically: $\delta\rho^{(0)}\propto {1}/{\sqrt{1-\xi^2}}$. It
describes charge distribution in the strip in response to a (uniform
along $x$ direction and smooth along $y$-direction) change of its
chemical potential~\cite{SE}. Other solutions of Eq.~(\ref{gagaps})
are found numerically:
 \bq\label{list}
\beta_1= 1.41,~~ \beta_2= 6.49,~~ \beta_3= 6.75, ...
 \ee
With increasing $n$ the eigenmodes of integro-differential equation
(\ref{gagaps}) oscillate faster, but in general do not follow the
oscillation theorem familiar from quantum mechanics. In particular,
 the solutions with $n=0$ and $n=3$ are even while $n=1$, $n=2$
are odd \cite{asymptotics}.


Finally, we mention the case of a gate-controlled $p$-$n$ junction,
Fig.~1b. The equilibrium density profile is linear near $x=0$ and
saturates for large $|x|$ \cite{ZF}. Eq.~(\ref{result}) is still
applicable for $q>1/d$. In the limit $q<1/d$ one should take into
account the screening of long-range Coulomb interaction by metallic
gates.
In this case the logarithm in the spectrum of the gapless plasmon
disappears, and the lowest mode Eq.~(\ref{gapless}) becomes
sound-like.


{\it Magnetoplasmons.} If external magnetic field ${\bf B}$ is
applied perpendicularly to the plane of graphene the plasmon spectra
acquire new modes. The equation of motion (\ref{motion}) should now
be modified to include the Lorentz force,
 \begin{equation}
\label{motion_Lorentz} {\bf \dot J}({\bf r},t) =\frac{e^2}{\pi
\hbar^2} |\mu(x)| {\bf E}({\bf r},t) -\frac{ev^2}{c \mu(x)}~ {\bf J}
\times {\bf B}.
 \end{equation}
The relative coefficient between electric and magnetic terms in this
equation  follows from the expression for the Lorentz force acting
on a single particle. The last term has opposite sign for electrons
and holes. Note that the frequency of cyclotron motion $\omega_B(x)
= e v^2B/c\mu(x)$ in graphene $p$-$n$ junctions is
position-dependent. The remaining equations
(\ref{continuity})-(\ref{coulomb}) are intact in the presence of
magnetic field. The boundary condition requires now the vanishing of
the normal component of electric current at the boundary, rather
than simply vanishing of the electric field, as in
Eq.~(\ref{boundary}). Eliminating ${\bf J}$ and ${\bf E}$ we arrive
at the generalization of equation (\ref{integro-diff}),
 \begin{eqnarray}
\label{with_marnetic_field} \delta \rho(x)+\frac{2e^2}{\pi}
\left\{q^2 {\cal Z}-\frac{q}{\omega} \left(\omega_B{\cal Z}\right)'
-\frac{d}{dx} {\cal Z} \frac{d}{dx} \right\} \nonumber \\
\times \int_{-d}^d d x' \delta \rho(x') K_0(|q||x-x'|)=0,~~~~
 \end{eqnarray}
where $ {\cal Z}(x) =|\mu(x)|/(\omega_B^2(x)-\omega^2).$

The most interesting effect described by
Eq.~(\ref{with_marnetic_field})  is the appearance of a set of new
modes, chiral magnetoplasmons, similar to those considered in
Ref.~\cite{AG} for conventional 2D electron systems with smooth
boundaries. To find their dispersion in strong magnetic fields, when
$\omega \ll \omega_B(x)$ (the exact condition is given below), one
should retain only the second term in
Eq.~(\ref{with_marnetic_field}). Noticing that $(\omega_BZ)'=\pi
l_B^2 \rho'_0(x)/e = \pi l_B^2/(l_E^2d)$, where $l_B=\sqrt{\hbar
c/eB}$ is the magnetic length, we arrive at the integral equation
\begin{eqnarray}
\label{strongB} -\frac{2 c}{ B}\frac{q}{ \omega}
\frac{d\rho_0(x)}{dx}\int_{-d}^d dx' \delta \rho(x')
K_0(|q||x-x'|)=\delta \rho(x).
\end{eqnarray}
Since $K_0$ is positive, propagation of magnetoplasmons with $q>0$
is quenched, indicative of their chiral property \cite{VM}. As seen
from Eq.~(\ref{strongB}), the plasmon  density $\delta\rho(x)$ is
concentrated where $\rho'_0(x)$ is the strongest. The derivative of
the charge density  in {\it field-induced} junctions
(\ref{approx_density}) features  strong singularity near the edges
of the flake. Thus, low-frequency magnetoplasmon spectrum is
strongly dependent on the microscopic regularization of this
singular behavior and is, therefore, beyond the scope of the
Thomas-Fermi approximation used throughout this paper.

The {\it gate-induced} junctions, however, allow a rather simple
analytical description of these modes if we approximate that
$\rho'_0(x)=e/l^2_Ed$ for $|x| \le d$ and $\rho'_0(x)=0$ for
$|x|>d$. The oscillating density $\delta \rho (x)$ then vanishes for
$|x|>d$. The solution inside the strip, $|x| \le d$, can be easily
found for $q \gg 1/d$, where one can assume the range of integration
in Eq.~(\ref{strongB}) to be infinite. The eigenfunctions of
Eq.~(\ref{strongB}) are simply given by $\sin{[q_\perp (x+d)]}$,
with the values of $q_\perp =\pi n/2d$ determined from the
condition, $\delta \rho_q (\pm d)=0$. The spectrum of
magnetoplasmons is then found to be,
\begin{equation}
\label{mp} \omega_n(q) = - \frac{2\pi e^2 l_B^2}{\hbar l_E^2 d}
\frac{q}{\sqrt{q^2+\pi^2 n^2/4d^2}},~~~n=1,2...
\end{equation}

The magnetoplasmon spectrum (\ref{mp}) is derived under the
assumption that magnetic field is strong, $\omega_B(d)\gg \omega$,
which implies that $l_B \ll l_E$. In order to neglect the first and
 third terms in the brackets in Eq.~(\ref{with_marnetic_field}) one
has to ensure that $q\ll (l_E/l_B)^4/d$. This condition might turn
out to be more or less restrictive than the hydrodynamics condition
$q\ll 1/l_E$, depending on the particular value of the ratio
$l_B/l_E$. Note that the smallness of this ratio is not in
contradiction to the non-quantized description of electron motion in
magnetic filed. The latter is valid as long as the filling factor is
large, $eEd \gg \omega_B(d)$, which means that $l_B\gg l_E^2/d$. For
magnetic field $\sim 1$$\text{T}$, and $l_B \sim 25$$\text{nm}$,
using the estimate below Eq.~(\ref{conditions}) that $l_E \sim
400$$\text{nm}$ we conclude that the width of the flake should
exceed $d>10\mu$$\text{m}$. The magnetoplasmon modes (\ref{mp}) are
$\sim (l_B/l_E)^2$ {\it slower} than electrons. Note that these
modes are undamped since single-particle excitations cannot be
induced at frequencies below cyclotron frequency $\omega_B$.

{\it Conclusions}. Graphene $p$-$n$ junctions are among the most
simple and  promising applications of this material. Single-electron
properties of $p$-$n$ junctions have been extensively studied. In
the present paper we investigated their collective excitations both
with and without magnetic field.
We anticipate that plasmon modes will be crucial for the optical
response of graphene nanostructures and realistic samples containing
electron-hole puddles. High degree of experimental control
should make them of special interest to nanoplasmonics and
electronics. Among the most promising applications of plasmons in
$p$-$n$ junctions we envisage a possibility of a ``plasmon
transistor'' \cite{HAA}. In particular, by simply switching the
direction of electric field from across the flake to along it (and
back) the propagation of plasmons can be facilitated (or prevented).
In addition, as follows from the above Eqs.~(\ref{result}),
(\ref{approx_density}), the plasmon velocity can be controlled with
simple change in the magnitude of  electric field. This is in a
sharp contrast to plasmons in metallic nanostructures, whose spectra
are typically fixed once devices are fabricated.


{\it Acknowledgments.} Useful discussions with M. Raikh and O.
Starykh are gratefully acknowledged. This work was supported by
DOE, 
Grant No.~DE-FG02-06ER46313. P.G.S. was supported by the SFB TR 12.


\begin{thebibliography}{50}
\bibitem[*]{} Present address: School of Physics, University of Exeter, EX4 4QL, U.K.
\bibitem{Wil} M. Wilson, Phys. Today {\bf 59}, No.~1, 21 (2006).

\bibitem{GN} A.K. Geim and K.S. Novoselov, Nature Mater. {\bf 6}, 183 (2007).

\bibitem{HAA} H.A. Atwater, Sci. Am. {\bf 296}, 56 (2007).

\bibitem{SAM} S.A. Maier, {\it Plasmonics: Fundamentals and Applications}
(Springer, New York, 2007).

\bibitem{Stern} F. Stern, Phys. Rev. Lett. {\bf 18}, 546
    (1967).

\bibitem{SFM} S. Gangadharaiah, A.M.  Farid, and E.G. Mishchenko,
Phys. Rev. Lett. {\bf 100}, 166802 (2008).

\bibitem{T_plasmon}  O. Vafek, Phys. Rev. Lett. {\bf 97}, 266406
(2006).

\bibitem{E_plasmon} E.H.~Hwang and S. Das Sarma,
Phys. Rev. B {\bf 75}, 205418 (2007).


\bibitem{cn} A.H.~Castro Neto {\it et al.},
 Rev. Mod. Phys. 81, 109 (2009).

\bibitem{puddles} J. Martin {\it et al.},
 Nature Physics {\bf 4}, 144 (2008).

\bibitem{Mar} J.R. Williams, L. DiCarlo, and C.M. Marcus, Science {\bf 317},
638 (2007).

\bibitem{LLV} Rigorous derivation of Eq.~(\ref{motion}) is
based on the ``relativistic'' stress energy-momentum tensor, see
L.D. Landau and E. M. Lifshitz, {\it Fluid Mechanics},
Butterworth-Heinemann, Oxford (1987), Ch.~15; M. Mueller, L. Fritz,
S. Sachdev, and J. Schmalian,  arXiv:0810.3657.

\bibitem{ftn1} In the case of gate controlled junctions the image
charges induced at the gates should be included into
Eq.~(\ref{coulomb}).

\bibitem{SMR}  T.A. Sedrakyan, E.G. Mishchenko, and  M.E. Raikh,
 Phys. Rev. B {\bf 74},
235423 (2006).


\bibitem{SE} P.G. Silvestrov and K.B. Efetov, Phys. Rev. B {\bf 77}, 155436
(2008).

\bibitem{asymptotics} In addition
even and odd solutions with $n>0$ have different singular behavior
at $|\xi|\ll 1$: $\delta\rho^{({\rm even})}\sim \sqrt{|\xi|}$,
$\delta\rho^{({\rm odd})}\sim {\rm sign}(\xi)/\sqrt{|\xi|}$. At
$\xi\rightarrow \pm 1$ all solutions diverge as $\delta\rho \sim
1/\sqrt{1-\xi^2}$.


\bibitem{ZF}  L.M. Zhang and M.M. Fogler, Phys. Rev. Lett. {\bf 100}, 116804
(2008).


\bibitem{AG} I.L. Aleiner and L.I. Glazman, Phys. Rev. Lett. {\bf 72}, 2935 (1994).

\bibitem{VM} V.A. Volkov and S. A. Mikhailov, JETP Lett. {\bf 42}, 556 (1985).





\end{thebibliography}
\end{document}